\documentclass[preprint,12pt]{elsarticle}
\usepackage{amsmath}

\usepackage{graphicx}
\usepackage{bm}
\usepackage{float}
\usepackage{mathptmx}
\usepackage{epstopdf}
\usepackage{booktabs}
\usepackage{multirow}
\usepackage{hhline}
\usepackage{verbatim}
\usepackage{subfigure}
\usepackage{cprotect}
\usepackage{lineno}
\usepackage{amssymb}
\usepackage{enumitem}   % % % added by 
\usepackage{soul}

\newcounter{bla}

\journal{Computer Physics Communications}

\begin{document}

\begin{frontmatter}


%\title{A \LaTeX{} template for CPC Computer Physics Descriptions}
\cprotect\title{\verb|AMMCR|: Ab-initio model for mobility and conductivity calculation by using Rode Algorithm}

\author[label1]{Anup Kumar Mandia}
%\author[label2]{Renuka Patnaik}
\author[label1]{Bhaskaran Muralidharan}
\author[label2]{Jung-Hae Choi}
\author[label3]{Seung-Cheol Lee\corref{author1}}
\author[label3]{Satadeep Bhattacharjee\corref{author2}}

\address[label1]{Indian Institute of Technology, Mumbai-400076, India}
\address[label2]{Center for Electronic Materials, Post-Silicon Semiconductor Institute, Korea Institute of Science and Technology, Seoul 02792, Republic of Korea}
\address[label3]{Indo-Korea Science and Technology Center, Bangalore 560065, India}

\cortext[author1] {Corresponding author.\\\textit{E-mail address:} seungcheol.lee@ikst.res.in}
\cortext[author2] {Corresponding author.\\\textit{E-mail address:} satadeep.bhattacharjee@ikst.res.in}

\begin{abstract}
We present a module to calculate the mobility and conductivity of semi-conducting materials using Rode's algorithm. This module uses a variety of electronic structure inputs derived from the Density Functional Theory (DFT). We have demonstrated good agreement with experimental results for the case of Cadmium Sulfide (CdS). We also provide a comparison with the widely used method, the so called Relaxation Time Approximation (RTA) and demonstrated the improvisation of the results compared to RTA. The present version of the  module is interfaced with Vienna \textit{ab-initio} simulation package (VASP). 

\end{abstract}


\end{frontmatter}

% Computer program descriptions should contain the following
% PROGRAM SUMMARY.

{\bf PROGRAM SUMMARY}
  %Delete as appropriate.

\begin{small}
\noindent
{\em Manuscript Title:}  \verb|AMMCR|: Ab-initio model for mobility and conductivity calculation by using Rode Algorithm \\
{\em Authors:}   Anup Kumar Mandia, Bhaskaran Muralidharan, Jung-Hae Choi, Seung-Cheol Lee and Satadeep Bhattacharjee                  \\
{\em Program Title:}   \verb|AMMCR|       \\
{\em Program obtainable from:}  www.ikst.res.in/ammcr\\
{\em Journal Reference:}                                      \\
  %Leave blank, supplied by Elsevier.
{\em Catalogue identifier:}                                   \\
  %Leave blank, supplied by Elsevier.
{\em Licensing provisions:} Open source BSD License       \\
  %enter "none" if CPC non-profit use license is sufficient.
{\em Programming language:}  C++                              \\
{\em Computer:} Any computer with GNU GCC compiler installed. Program is tested with gcc version 5.4.0.      \\
  %Computer(s) for which program has been designed.
{\em Operating system:} Unix/Linux                    \\
  %Operating system(s) for which program has been designed.
  % Fill in if necessary, otherwise leave out.
  % Please give some freely chosen keywords that we can use in a
  % Fill in if necessary, otherwise leave out.
%{\em Subprograms used:}                                       \\
  %Fill in if necessary, otherwise leave out.
{\em Nature of problem:} Long-range interactions between electrons and longitudinal optical (LO) phonons are a major challenge in computing transport in polar semiconductors. Due to such LO phonons or generally polar optical phonons (POP), electron phonon scattering requires special attention since they can not be studied using a simple frame work such as relaxation time approximation (RTA).  \\ %Describe the nature of the problem here.
{\em Solution method:} We have developed a code that calculates mobility and conductivity by using \textit{abinitio} inputs with Rode algorithm which treats the interaction between the polar optical phonons and electrons in a proper way. \\
  %Describe the method solution here.
%   \\
%{\em Reasons for the new version:}*\\
  %Only required for a New Version summary, otherwise leave out.
%   \\
%{\em Restrictions:}\\
  %Describe any restrictions on the complexity of the problem here.
%   \\
%{\em Unusual features:}\\
  %Describe any unusual features of the program/problem here.
%   \\
%{\em Additional comments:}\\
  %Provide any additional comments here.
%   \\
%#{\em Running time:} 1-600 minutes (depends on the number of processors and system size)\\
  %Give an indication of the typical running time here.
%   \\
\end{small}

\section{Introduction}
With the ongoing development in the field of density functional theory (DFT), tremendous progress has been made in obtaining details in the electronic structure of complex materials important for understanding the physical properties of the materials in the ground state as well as in the excited state. This resulted in many computational tools which rely on first-principles approach based on DFT. To name a few: Vienna Ab initio Simulation Package (VASP) \cite{vasp1}-\cite{vasp5}, Quantum Espresso \cite{qe}, Spanish Initiative for Electronic Simulations with Thousands of Atoms (SIESTA) \cite{siesta1,siesta2} etc. The use of DFT obtained results as inputs for the study of the non-equilibrium properties such as transport is another area which is developing rapidly for a decade or so. Methods are being developed to study the transport properties at various levels: From semi-classical Boltzmann level to Green-Kubo formalism, Landauer formalism, non-equilibrium Green's function formalism etc \cite{T1,T1-1,T2,T3}. There are few semi classical method based on Boltzmann Transport equation (BTE) \cite{rode1}-\cite{series_exp2} for the calculation of transport coefficients  which use inputs obtained from either first-principles based calculations or experiments. For example, the code \verb|BoltzTrap| \cite{boltztrap} was developed in a way that it can accept the \textit{abinitio} inputs from the DFT-based calculations and is widely used in scientific community. \verb|BoltzTrap| is successful in studying the transport behaviour in variety of materials ranging from conductors, inter-metallics to thermoelectrics. One limitation in such formalism is the use of so called constant Relaxation time approximation (c-RTA). It is well known that Relaxation time approximation (RTA) is not  appropriate for inelastic scattering mechanisms \cite{lundstrom}. The use of c-RTA even more simplifies the problem by assuming a universal constant relaxation time for all elastic and inelastic scattering processes. 
\par
In the recent years, there has been an enhanced focus on the next-generation semiconductors made from two or more elements,  particularly one that is strongly doped or highly degenerated. In specific, the so-called III-V materials are of present attention due to their very high carrier mobility compared to silicon. One important point about these compound semiconductors is that, unlike silicon, the carrier scattering due to the polar optical phonons is very essential, especially the temperature at which the device would be operational. These scattering processes are inelastic and one therefore requires a model for carrier transport that can take proper account of such inelastic scattering from a computational point of perspective.
\par
In the present work, we have developed a code which is based on the Rode iterative method \cite{rode1,rode2,rode3} and uses the \textit{ab-initio} inputs and at present interfaced with the  $\mathrm{VASP}$ code. Using such approach in recent studies, we have demonstrated the effect of inelastic scattering on the transport properties of n-type ZnSe~\cite{Anup} and AlGaAs$_2$ ~\cite{Soubhik}.
The code uses electronic band structure, density of states, phonon frequencies, dielectric, piezoelectric and elastic tensors as inputs for the simulation. The band structure and density of states  must be in $\mathrm{VASP}$ output format, while the other  inputs such as the dielectric, elastic and piezoelectric constants could be either from \textit{abinitio} calculations or from the experimental results depending on the choice of the user.
\par
In this manuscript, we present a general implementation of \verb|AMMCR| module which can be easily interfaced with VASP tool. The module \verb|AMMCR| is written in C++. In this module eight types of scattering mechanisms are included, these are ionized impurity scattering, polar optical phonon scattering due to longitudinal phonon, acoustic deformation scattering, piezoelectric scattering, dislocation scattering, alloy scattering, intra-valley scattering and neutral impurity scattering. Out of these eight scattering mechanism any scattering can be included or excluded from simulation. We have tested \verb|AMMCR| for CdS and found reasonable agreement with the experiment. We have also demonstrated that results with the Rode iterative method is much better than relaxation time approximation (RTA). \\
This paper is organized as follows: section 2 describes the the Boltzmann transport formalism, Rode algorithm and different scattering mechanisms; section 3 describes the structure of the code; section 4 describes the input files required for simulation; test system for Cadmium Sulfide (CdS) is discussed in section 5 and we conclude in section 6.

\section{Theoretical framework}
\subsection{Solution of  Boltzmann Transport Equation via iterative method}
The motion of the carriers are described via a  probability distribution function \textit{f(\textbf{r},\textbf{k},t)} in real and momentum space and as a function of time. \textit{f(\textbf{r},\textbf{k},t)} is the solution of the Boltzmann transport equation given by \cite{lundstrom,jasp,ferry}
\begin{equation}
\frac{\partial \textit{f}}{\partial t} + \textbf{v}\cdot\nabla _rf + \frac{e\textbf{E}}{\hbar}\cdot\nabla _kf= \frac{\partial \textit{f}}{\partial t}\Bigr|_{\substack{coll}}
\label{BTEE}
\end{equation}
Where \textbf{v} is the group velocity of the carriers, \textbf{E} is the applied electric field, $\frac{\partial \textit{f}}{\partial t}\Bigr|_{\substack{coll}}$ represents the change in the distribution function with time due to collisions. 

For steady state, $\frac{\partial \textit{f}}{\partial t}=0$, and assuming spatial homogeneity ($\nabla _rf=0)$, the Eq.\ref{BTEE} can be written as
\begin{equation}
\frac{e\textbf{E}}{\hbar}\cdot\nabla _kf
=\int dk'\lbrace s(k',k)f(k)(1-f(k))-s(k,k')f(k)(1-f(k'))\rbrace
\label{BTEF}
\end{equation}
$S(k',k)$ represents scattering rate for an electron making a transition from a state $k$ to another
state $k'$.
In the presence of low electric field the distribution function is written as 
\begin{equation} 
f(k) = f_0[\epsilon(k)] +  g(k)cos\theta,
\label{distft}
\end{equation}
Where $f_0[\epsilon(k)]$ is the equilibrium distribution function, and \(\cos\theta\) is the angle between applied electric field and \(k\). $g(k)$ is the perturbation to the equilibrium distribution function. Higher order terms are neglected here, since transport coefficients are calculated under low electric field conditions. Now, we have to calculate the perturbation $g(k)$ to calculate transport coefficients. The perturbation in distribution function can be obtained by inserting the $f(k)$ and $f(k')$ from the Eq.\ref{distft} to the Eq.\ref{BTEF}. After some manipulation, $g(k)$ can found as
\cite{rode1,rode2,rode3,Anup,Soubhik}
\begin{equation}
g_{k,i+1} = \frac{S_i(g_k,i)- \frac{eE}{\hbar}(\frac{\partial f}{\partial k}) } {S_o(k)+ \frac{1}{\tau_{el}(k)}} .
\label{pert}
\end{equation}
As $g_{k}$ appears on both the left and right hand sides of the Eq.\ref{pert}, the above equation should be solved in iterative manner. Here i is the iteration index. $\frac{1}{\tau_{el}(k)}$ is the sum of momentum relaxation rates of all elastic scattering rates,  $S_{i}$ and $S_{o}$ are in scattering and out scattering operator for inelastic scattering and are given by \cite{rode3}. 
\begin{equation}
\ S_o(k) = \int [s(k,k')(1 - f') + s(k',k)f']dk'
\label{outsc}
\end{equation}
\begin{equation}
\ S_i(g_k,i) = \int X g_{k',i} [s(k',k)(1-f) + s(k,k')f ]dk'
\label{insc}
\end{equation}
where X is the cosine of the angle between the initial and the final wave vectors and $f'=f(k')$.   Since convergence is exponential \cite{rode3}, it requires only a few iterations for convergence. $\frac{1}{\tau_{el}(k)}$ is given by 
\begin{equation}
\ \frac{1}{\tau_{el}(k)} = \frac{1}{\tau_{ii}(k)} + \frac{1}{\tau_{ac}(k)} + \frac{1}{\tau_{pz}(k)} + \frac{1}{\tau_{dis}(k)} + \frac{1}{\tau_{alloy}(k)} + \frac{1}{\tau_{iv}(k)} + \frac{1}{\tau_{ni}(k)}
\label{elst}
\end{equation}
where the subscripts \textit{el},\textit{ii}, \textit{ac}, \textit{pz}, \textit{dis}, \textit{alloy}, \textit{iv} and \textit{ni} are stands for elastic, ionized impurity, acoustic deformation potential, piezoelectric, dislocation, alloy, intra-valley and neutral impurity scattering processes respectively. For calculating mobility, thermal driving force, $v(\frac{\partial f}{\partial z})$ in Eq. \ref{pert} is set to zero, only electric driving force $\frac{eE}{\hbar}(\frac{\partial f}{\partial k})$ is considered. The carrier mobility $\mu$  is  given by \cite{rode1,rode2,rode3,alireza}
\begin{equation}
\mu = \frac{1}{3E} \frac{\int v(\epsilon) D_s(\epsilon) g(\epsilon) d\epsilon}  
{\int D_s(\epsilon) f( \epsilon )  d\epsilon },
\label{mobility}
\end{equation}
\quad
where $D_S(\epsilon)$ represents density of states. The carrier velocity $v(k)$ is calculated from abinito band structure by using relation
\begin{equation}
\ v(k) = \frac{1}{\hbar} \frac{\partial{\epsilon}}{\partial{k}}.
\label{velocity}
\end{equation}
After calculating mobility electrical conductivity can be calculated by 
\begin{equation}
\ \sigma = n e \mu_e,
\label{conductivity}
\end{equation}
where $n$ is the electron carrier concentration, and $\mu_e$ is the electron mobility. 

\subsection{Scattering Mechanisms}
Eight different types of scattering mechanisms are included in this code. Out of these eight scattering mechanisms any scattering mechanism can be included in the simulation. These eight scattering mechanisms are ionized impurity scattering, Polar Optical phonon scattering due to longitudinal phonon, acoustic deformation scattering, piezoelectric scattering, dislocation scattering, alloy scattering, intra-valley scattering and neutral impurity scattering. 
\newline Interaction of electrons with ionized impurity potential results in ionized impurity scattering. Ionized impurity scattering is an important scattering mechanism at higher doping concentration and at lower temperatures. For ionized impurity scattering Brooks-Herring approach \cite{brook} is used. The momentum relaxation rate for ionized impurity scattering is given by \cite{rode3,alireza}
\begin{equation}
\frac{1}{\tau_{ii}(k)} = \frac{e^4 N }{8\pi\epsilon_0^2\hbar^2 k^2 v(k)}[D(k)ln(1+\frac{4k^2}{\beta^2})-B(k)] ,
\label{Ionized impurity}
\end{equation}
where $\epsilon_0$ is the dielectric constant and $\beta$ is the inverse screening length given by
\begin{equation}
\beta^2 = \frac{e^2}{\epsilon_0 k_B T}\int D_s(\epsilon) f(1-f)d\epsilon,
\label{beta square}
\end{equation}
where N is the concentration of ionized impurity and it is given by
\begin{equation}
\ N = N_A + N_D
\label{impurity}
\end{equation} 
where $N_A$ and $N_D$ are the acceptor and donor concentrations respectively. The expressions for D(k) and B(k) are \cite{rode3}. \\
\begin{equation}
\ D(k) = 1 + \frac{2 \beta^2 c^2}{k^2} + \frac{3 \beta^4 c^4}{4 k^4}
\label{D(k)}
\end{equation} 

\begin{equation}
\ B(k) = \frac{4 k^2/\beta^2}{1 + 4 k^2/\beta^2} + 8 \frac{\beta^2+2k^2}{\beta^2+4k^2}c^2 +   \frac{3\beta^4+6\beta^2 k^2 -8k^4}{(\beta^2+4k^2)k^2}c^4
\label{B(k)}
\end{equation} 
The wave function admixture $c(k)$ is the contribution of p-orbital to the wave function of the band. It is calculated by projecting the Kohn-Sham wavefunctions onto the spherical harmonics which are non-zero only within the spheres centering the ions and this is already implemented in \textsl{VASP} package.

Polar optical phonon (POP) scattering is inelastic and an-isotropic scattering mechanism. In most of the polar semiconductors POP scattering is dominant scattering mechanism near room temperature or in the higher temperature region. The \textit{out scattering} operator is given by \cite{rode3}
\begin{equation}
S_{o} = ( N_{po} + 1 - \textit{f}^- ) \lambda^{-}_0 + ( N_{po} + \textit{f}^+) \lambda^{+}_0 
\label{So}
\end{equation}

\begin{equation}
\lambda^+_{o} = \beta^+[(A^+)^2 ln \mid \frac{k^+ + k}{k^+ - k}\mid - A^+ c c^+ - aa^+cc^+] 
\label{lambda0}
\end{equation}

\begin{equation}
\beta^+ = \frac{e^2 \omega_{po} k^+ }{4 \pi \hbar k v(k^+)} (\frac{1}{\epsilon_\infty} - \frac{1}{\epsilon_0})
\label{betaaaa}
\end{equation}

\begin{equation}
A^+ = aa^+ + \frac{(k^+)^2 + k^2}{2 k^+ k} cc^+,
\label{Apositive}
\end{equation}

Here, a(k) gives the contribution of s-electrons to the band, and in a way similar to c(k) it is calculated from the first-principles. $k^{\pm}$ is the solution of the equation, $\varepsilon (k) \pm \hbar \omega_{po}$. Any quantity with subscript plus/minus has to be evaluated at energy corresponding to k$^+$/k$^-$. $\lambda_o$ are rate of \textit{out} scattering. The subscript plus and minus denotes the scattering by the absorption or emission respectively, so it is to be evaluated at an energy $\epsilon + \hbar \omega_{po} $ for absorption and at energy  $\varepsilon(k) - \hbar \omega_{po} $ for emission. If the energy of phonons is less than $ \hbar \omega_{po} $, the emission of phonons is not possible and hence $\lambda^{-}_o$ is to be considered to be zero. \\ 
$N_{po} $ is the number of phonons and is given by \cite{rode3}
\begin{equation}
N_{po} = \frac{1}{exp(\hbar \omega_{po}/k_B T)-1} .
\label{Npo}
\end{equation}
The {scattering in} operator $S_{i}$ is given by \cite{rode3}
\begin{equation}
S_{i} = ( N_{po} + \textit{f} ) \lambda^{-}_i g^- + ( N_{po} + 1 - \textit{f}) \lambda^{+}_i g^+ 
\label{Si}
\end{equation}
Here $\lambda_i$ is the rate of \textit{in scattering}. Again, the plus and minus subscriptions indicate absorption and emission, respectively.
\begin{equation}
\lambda^+_{i}(k) = \beta^+ [\frac{(k^+)^2+k^2}{2k^+k}(A^+)^2 ln\mid \frac{k^+ + k}{k^+ - k}\mid - (A^+)^2 - \frac{c^2 (c^+)^2}{3}]
\label{lambdai}
\end{equation}

Coupling of electrons with non-polar acoustic phonons results in acoustic deformation potential scattering. The momentum relaxation rate for acoustic deformation potential scattering is given by \cite{rode3,alireza}
\begin{equation}
\frac{1}{\tau_{ac}(k)} = \frac{e^2 k_B T E_D^2 k^2}{3\pi\hbar^2 c_{el}v(k)}[3 - 8c^2(k)+6c^4(k)] ,
\label{acoustic deformation}
\end{equation}
Where c$_{el}$ is the spherically averaged elastic constant.
$E_D$ is acoustic deformation potential and is given by conduction band shift (in eV) per unit strain due to acoustic waves. \\
Polar scattering due to acoustic phonons is called piezoelectric scattering. Piezoelectric scattering is important at low doping concentration and at low temperature in polar materials. The momentum relaxation rate for piezoelectric scattering with ab-initio parameters is given by \cite{rode3, alireza}
\begin{equation}
\frac{1}{\tau_{pz}(k)} = \frac{e^2 k_B T P^2 }{6\pi\epsilon_0\hbar^2 v(k)}[3 - 6c^2(k)+4c^4(k)] 
\label{pz}
\end{equation}
where $P$ is a dimensionless piezoelectric coefficient, it is isotropic for zinc blende structure and anisotropic for wurtzite structure (see the appendix). 
\quad
\newline The momentum relaxation rate for dislocation scattering is given by \cite{miller}
\begin{equation}
\frac{1}{\tau_{dis}(k)} = \frac{N_{dis}e^4k}{\hbar^2 \epsilon_0^2 c_l^2v(k)} \frac{1}{(1+\frac{4k^2}{\beta^2})^{3/2}\beta^4}
\label{dislocation}
\end{equation}
In case of alloy there is one more scattering mechanism due to atomic disorder, known as alloy scattering. The momentum relaxation rate for alloy scattering is given by \cite{ramu}
\begin{equation}
\frac{1}{\tau_{alloy}(k)} = \frac{3 \pi k^2}{16 \hbar^2 v(k)} V_0 U_{alloy}^2 \chi (1-\chi)
\label{alloy}
\end{equation}
The momentum relaxation rate for intra-valley scattering is given by \cite{rode3}
\begin{equation}
\frac{1}{\tau_{iv}(k)} = (N_{e} + 1 - f^-) \lambda_e^- + (N_{e} + f^+) \lambda_e^+ 
\label{iv}
\end{equation}
where $N_e$ is the phonon occupation number and given by
\begin{equation}
N_e = \frac{1}{exp(\frac{\hbar \omega_e}{k_B T})-1} 
\label{N_e}
\end{equation}  
$\lambda_e^+$ is given by
\begin{equation}
\lambda_e^+ = \frac{e^2D_e^2(Z-1)kk^+}{2 \pi \rho \hbar v(k) \omega_e} 
\label{lambda_e_pos}
\end{equation}  
Similarly $\lambda_e^-$ is given by
\begin{equation}
\lambda_e^- = \frac{e^2D_e^2 Z kk^-}{2 \pi \rho \hbar v(k) \omega_e} 
\label{lambda_e_pos}
\end{equation}  
where $Z$ is the number of equivalent valleys, $D_e$ is the intervalley deformation potential (units of electron volts per meter), $\hbar \omega_e$ is the phonon energy, $\rho$ is the density of the material. If $E < \hbar \omega$,  $\lambda_e^-$ is considered to zero.\\ 
There is one impurity scattering due to non ionized donors called neutral impurity scattering. Neutral impurity scattering is important at higher doping concentration and low temperature. For neutral impurity scattering we have used Erginsoy model\cite{erginsoy}. The momentum relaxation rate of neutral impurity scattering is given by
\begin{equation}
\frac{1}{\tau_{ni}} = \frac{80 \pi \epsilon \hbar v(k)^2 N_n}{e^2 k^2}
\label{neutral_impurity}
\end{equation}  
where $N_n$ is concentration of neutral impurities in semiconductor.  
\subsection{Ab-initio Inputs}
For calculating transport properties, band structure and density of states of semiconductor material is required. The required band structure and density of states are calculated by using the density functional theory (DFT) using a three dimensional \textit{k} mesh around the conduction band minimum (CBM). Near CBM much finer meshing is required to obtain good results, since this near region to CBM will play a major role at low electric field. To obtain a k-point file in reciprocal space for DFT calculation a program k\_point\_generator.cpp is given with the code. The program is saved in the \textquoteleft utility\textquoteright{} folder. The conduction band is represented by a function of distance from the CBM, by taking the average  of the energy values of the k-points that are at the same distance from CBM. We have done an analytical fitting of the band with a six degree polynomial, to get smooth curves for group velocity. The group velocity is calculated by using equation
\begin{equation}
v(k) = \frac{1}{\hbar} \frac{\partial \epsilon(k)}{\partial k}
\label{neutral_impurity}
\end{equation}  
The calculated group velocity by the above equation is used to calculate required different scattering rates. All other inputs required for calculations, low and high frequency dielectric constant \cite{low,high}, elastic constant, piezoelectric constant and polar optical phonon frequency $\omega_{po}$ \cite{polar} are calculated by using Density functional theory (DFT) by using VASP. The acoustic deformation potential is given by equation
\begin{equation}
E_D = -V (\frac{\partial E_{CBM}}{\partial V})\Bigr|_{V=V_0}
\end{equation}
where V is the volume, $V_0$ is the volume under zero pressure and CBM is conduction band minima \cite{acoustic1,acoustic2}. So, all required inputs are calculated by using DFT. Only crystal structure of semiconductor material is given as input. So, there is no need to heavily depend on experimental data for mobility and conductivity calculation. 
\subsection{Simulation Flowchart}
\begin{figure}
\centering
\includegraphics[width=180mm,height=200mm]{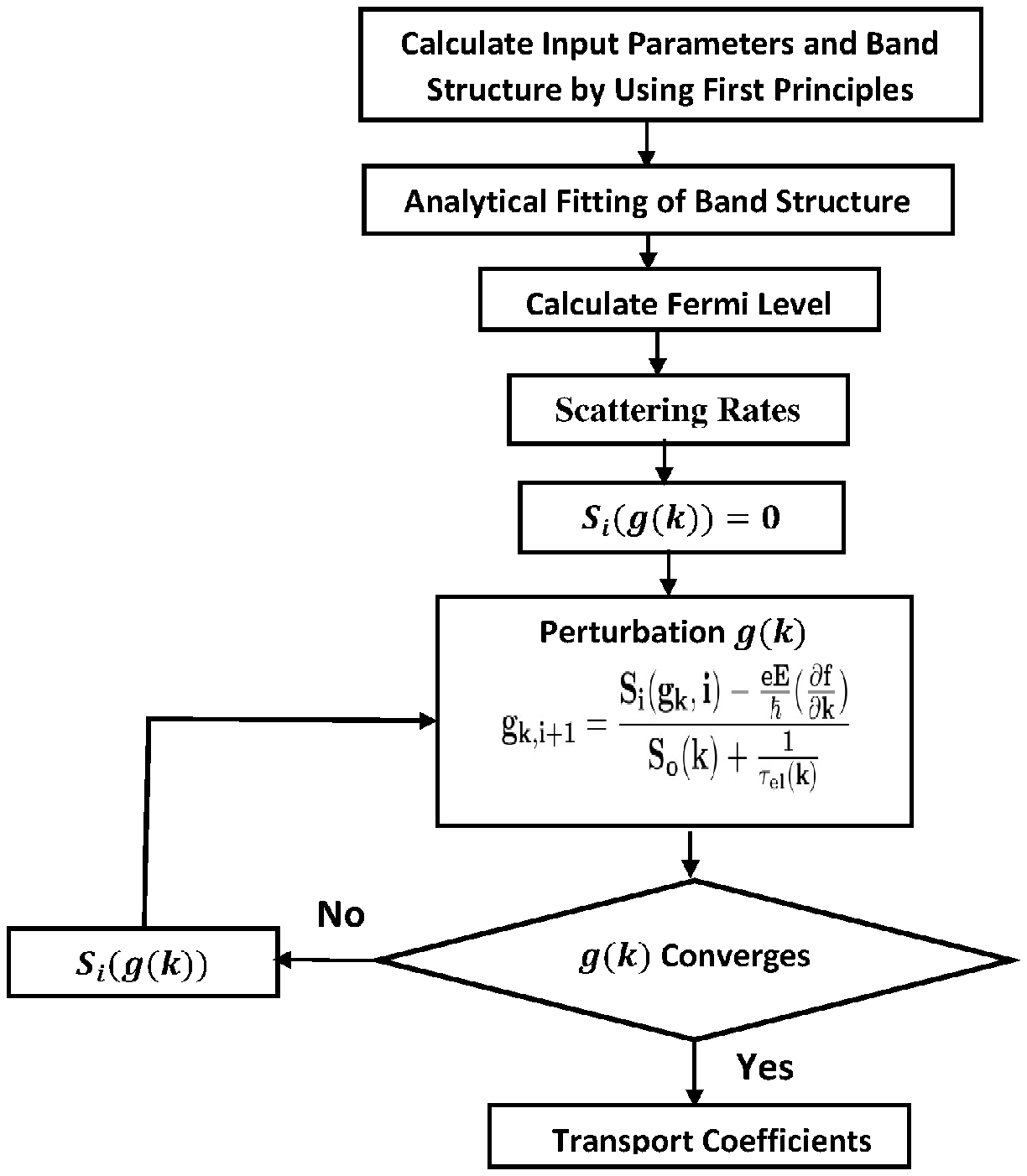}
\caption{ Flowchart for transport  calculation from \textit{ab-initio} inputs}
\label{flowchart}
\end{figure}

Fig. \ref{flowchart} shows the step to calculate mobility and conductivity using \verb|AMMCR| code. First, we have to calculate band structure and all other input parameters as explained in the previous section by using first principles methods. Then analytical fitting of band structure is done, to obtain a smooth curve for group velocity. The Fermi level is calculated with smooth band structure obtained after analytical fitting by using equation,
\begin{equation}
n = \frac{1}{V_0} \int_{\epsilon_c}^{\infty}{D_S(\epsilon)f(\epsilon) d\epsilon},
\label{fermi}
\end{equation}
where $D_S(\epsilon)$ represents the density of states at energy $\epsilon$, where $\epsilon_c$ represents the bottom of the conduction band and $V_0$ represents the volume of the cell. Next, scattering rates are being calculated for different selected scattering mechanisms by using Eq.s \ref{Ionized impurity}, \ref{So}, \ref{acoustic deformation}, \ref{pz}, \ref{dislocation}, \ref{alloy}, \ref{iv}, \ref{neutral_impurity}. The perturbation in distribution function $g(k)$ is calculated by using Eq.\ref{pert}, keeping  $S_i(k)=0$, this particular step gives the results within RTA. To obtain the results beyond RTA, $g(k)$ in is calculated by iteration till $g(k)$ (or $g(\epsilon)$) converges. The $g(\epsilon)$ obtained in the above manner is used to calculate the transport coefficients. 
\section{Code Layout}
Upon unzipping the tar file, the main folder "AMMCR" contains four sub-folders: \textquoteleft src\textquoteright{}, \textquoteleft utility\textquoteright{},\textquoteleft example \textquoteright{} and  \textquoteleft manual\textquoteright{}. \textquoteleft src\textquoteright{} folder contains different functions to calculate mobility and conductivity. It contains a file \textquoteleft main.cpp\textquoteright{}, it is the main file, that calls all other functions for calculation. \textquoteleft utility\textquoteright{} folder contains a file \textquoteleft k\_point\_generator.cpp\textquoteright{}, this file contains a program that will generate a \textquoteleft k\_points\_file \textquoteright{} file, containing the k-points required for the DFT simulation of the band structure using VASP. \textquoteleft manual\textquoteright{} contains the user-guide while the folder \textquoteleft example \textquoteright{} contains the example files.

\section{Input Files}
The execution of \verb|AMMCR| module requires five files as input. Four input files EIGENVAL, OUTCAR, PROCAR and DOSCAR are to be obtained by using VASP package and one input file input.dat that contains the value of different material constants calculated by abinito method as explained in section 2.3. Sample input files are given in the \textquoteleft example\textquoteright{} folder. 
Typical inputs for a given donor concentration is given below (for the case of CdS),
\begin{verbatim}
32.75 37.75 48.43 61.05 75.55 99.12 123.99 154.08 154.13 171.998 194.67 264.20 299.94 346.74 411.18    # Temperature loop
6.9e15					  # Donner conc in cm^-3  	
0						  # Acceptor conc in cm^-3 	
0						  # Neutral impurity conc cm^-3 	
9.97			          # Static dielectric const
6.1						  # High frequency dielectric const	
0						  # Band gap (eV) 
1						  # Equivalent number of valence band valleys	
1						  # Equivalent number of conduction band valleys
0						  # Density in gm/cm^3
1 1 1 1 0 0 0 0		      # Different scatt mechanism included
0						  # Dislocation density cm^-2  
6.44					  # Longitudinal Optical freq in THZ	
12						  # Acoustic deformation potential in eV
0.141104752018947 		  # Piezo Electric coefficient (dimensional less)
7.6924e+11  		      # Longitudianl elastic constant in dyne/cm^2
1.9146e+11				  # Transverse elastic constant in dyne/cm^2
0						  # Alloy potential in eV
0						  # Volume of unit cell in nm^-3
0						  # Fraction of atom for alloy
0						  # Phonon freq for intravalley scatt in THz 
0						  # Coupling constant for intravalley scattering in 10^8 eV/cm 
0						  # Number of final valley for intravalley scattering 
0						  # Density of state 0 read from DOSCAR; 1 for free electron density
10						  # Maximum number of iteration
\end{verbatim}

\section{Test System} % Example 
\begin{figure}
\includegraphics[width=125mm,height=100mm]{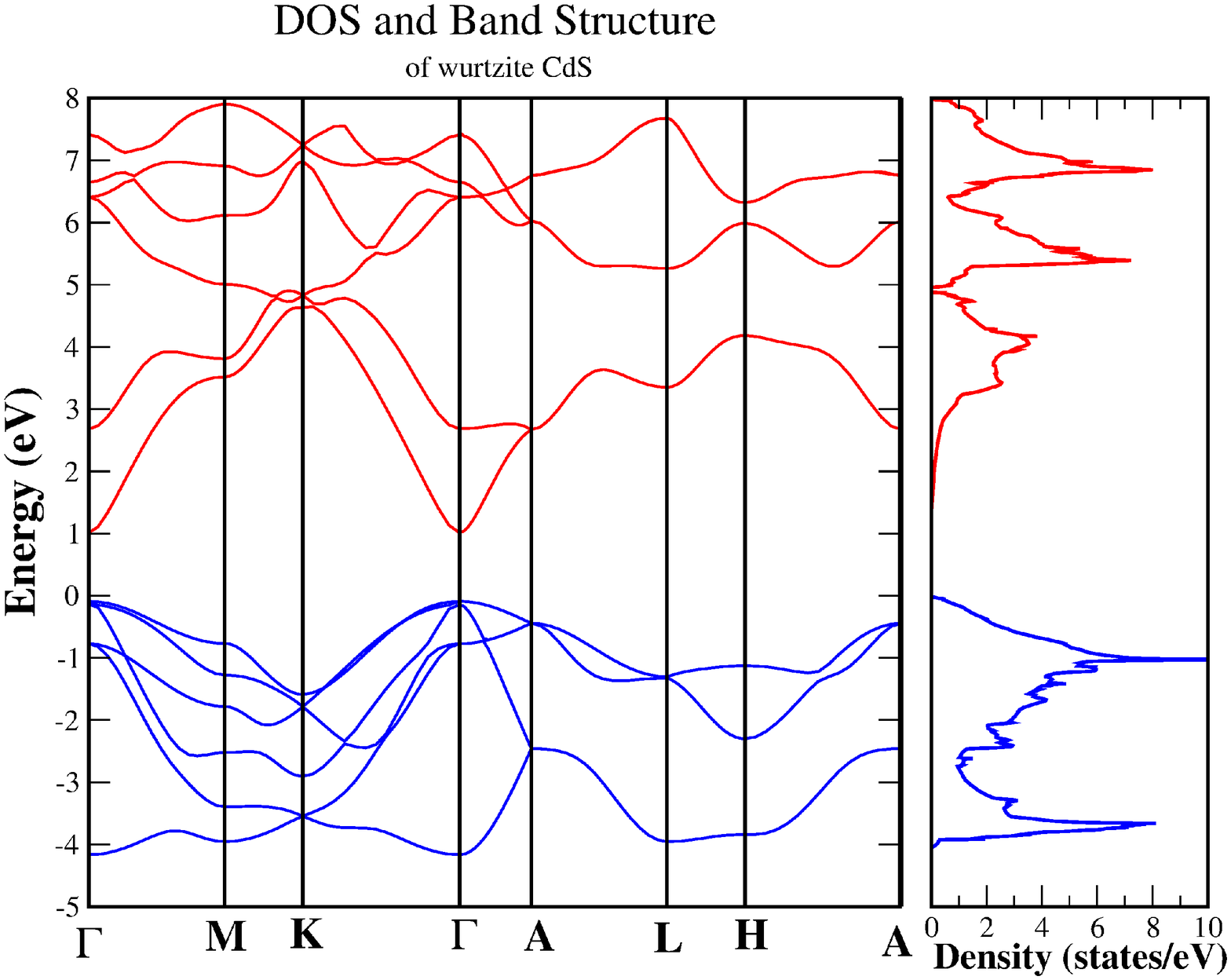}
\caption{Band Structure and Density of states for Wurtzite CdS}
\label{band}
\end{figure}

We have calculated the transport properties of bulk CdS using the \verb|AMMCR| code. We have considered the wurtzite structure of CdS (space group No.:186). The electronic structure calculations were performed within the frame-work of density functional theory (DFT) with Perdew-Burke Ernzerhof exchange correlation energy functional which is based on a generalized gradient approximation. Fig. \ref{band} shows electronic band structure and density of states (DOS) obtained for CdS.
We have calculated all required input parameters by using abinito principle. As an input for the transport calculation within Rode’s method, only band structure for one valley is needed, we have therefore performed non-self consistent calculations of the band energies in a
special k-point mesh around the $\mathrm{\Gamma}$ point with 8531 k-points. Such an approach enables us to efficiently account for the group velocity. 
As CdS has wurtzite structure, it has different dielectric permittivity in parallel and perpendicular to the c axis. For calculation isotropic value of permittivity given by equation
\begin{equation}
\epsilon = \frac{1}{3}(\epsilon_{\parallel} + 2 \epsilon_{\perp})
\end{equation} 
is used. Calculated value of low and high frequency dielectric constant are 9.97 and 6.1 respectively. We have obtained a \textit{abinitio} value of polar optical phonon frequency of 6.44 THz. We have obtained abinito value of lattice constant $a_0 = 0.420 nm$ and $c_0 = 0.687 nm$, 12 eV acoustic deformation potential, 0.10059 and  0.141104 of perpendicular and parallel piezoelectric coefficient. Calculated value of elastic constant are $c_{11}=7.77 \times 10^{10} N/m^2$ , $c_{13}=3.72 \times 10^{10} N/m^2$, $c_{33}= 8.79 \times 10^{10} N/m^2$ and $c_{44} = 1.49 \times 10^{10}  N/m^2$ respectively.       

\begin{figure}
%\centering
%\subfigure[ $$ 30 K]{\includegraphics[width=100mm,height=80mm]{scattering_rate_30_1e13.eps} }
%\subfigure[ $$ 300 K]{\includegraphics[width=100mm,height=80mm]{scattering_rate_300_1e13.eps} }
\includegraphics[width=120mm,height=140mm]{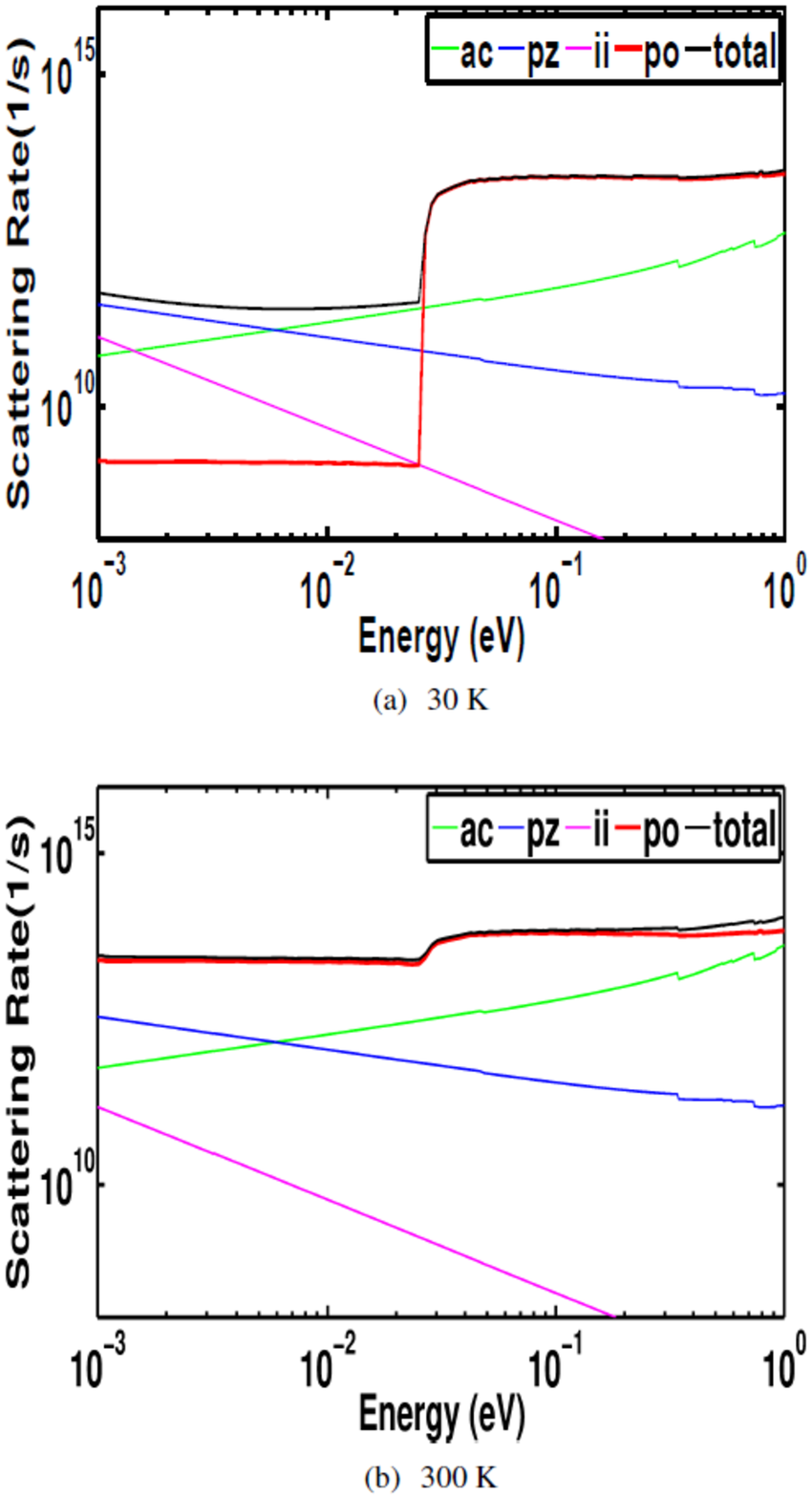}
\caption{Scattering Rates vs Energy for $N_D=1\times10^{13}cm^{-3}$}
\label{scattering_13}
\end{figure}

\begin{figure}
%\centering
%\subfigure[ $$ 30 K]{\includegraphics[width=100mm,height=80mm]{scattering_rate_30_1e15.eps} }
%\subfigure[ $$ 300 K]{\includegraphics[width=100mm,height=80mm]{scattering_rate_300_1e15.eps} }
\includegraphics[width=120mm,height=140mm]{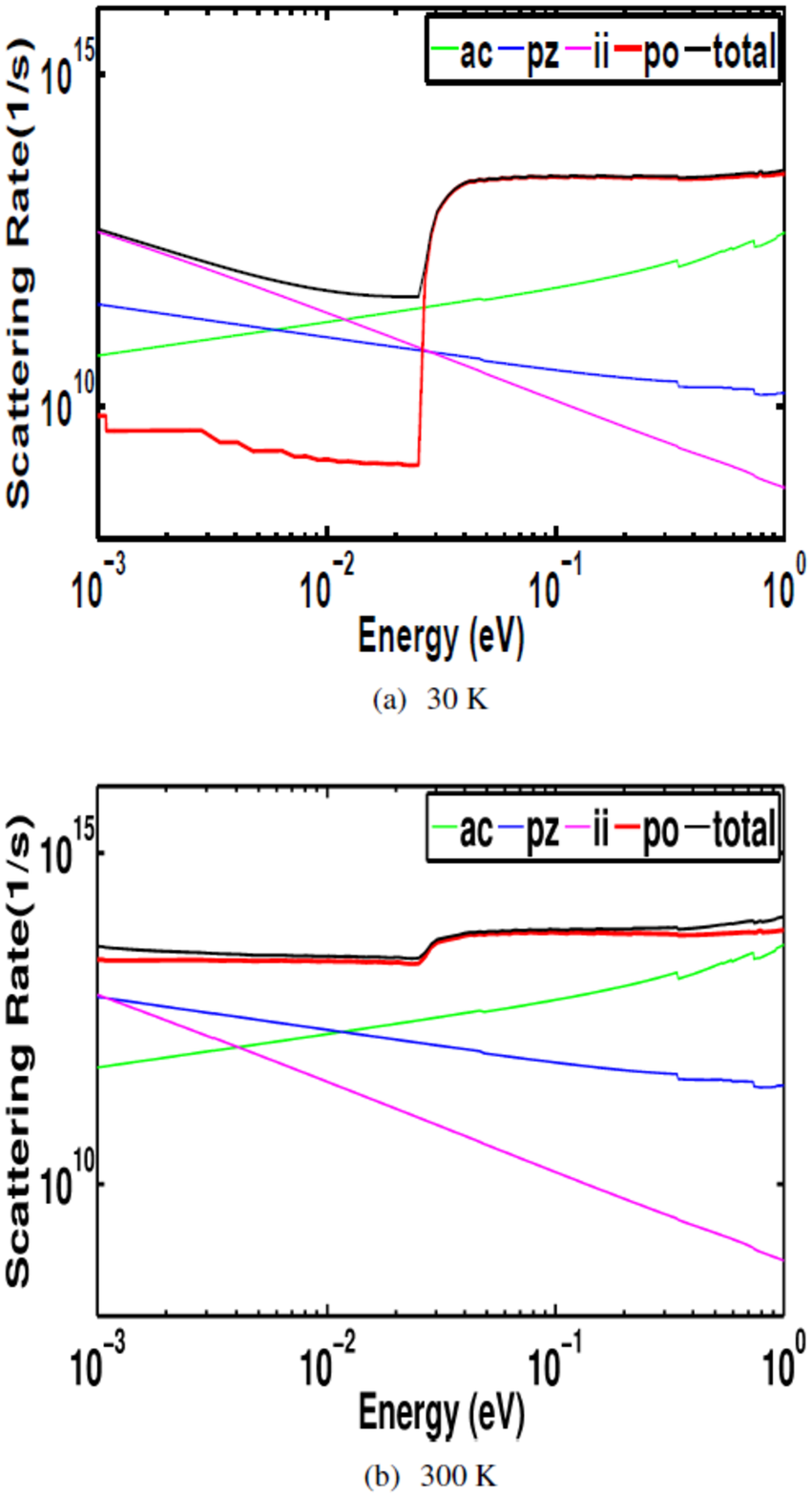}
\caption{Scattering Rates vs Energy for $N_D=1\times10^{15}cm^{-3}$}
\label{scattering_15}
\end{figure}

\begin{figure}
\centering
\includegraphics[width=125mm,height=100mm]{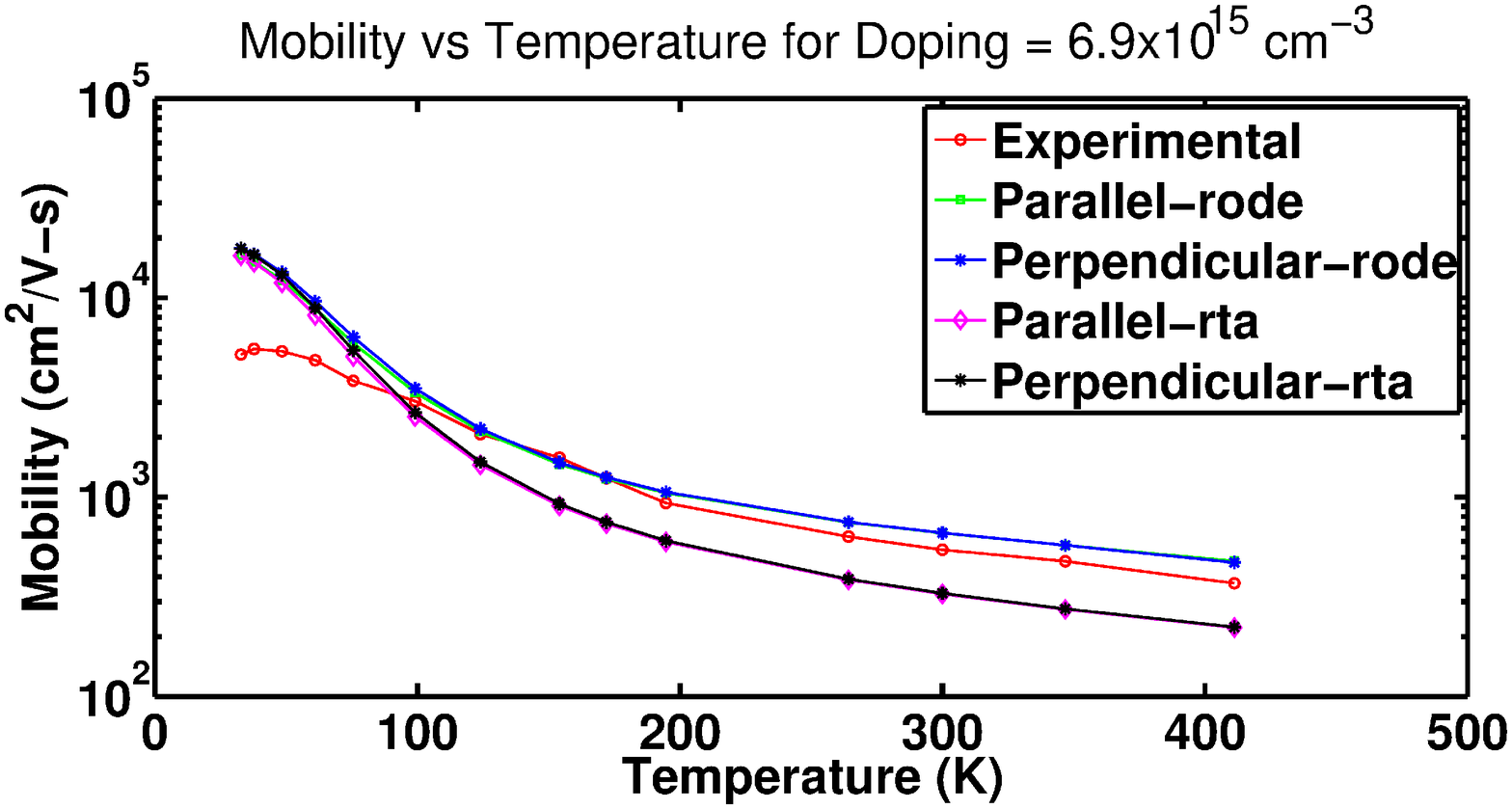}
\caption{Calculated and experimental measured mobility with temperature variation for CdS at Doping  $6.9 \times 10^{15} cm^{-3}$ \cite{crandall}.}
\label{m-t}
\end{figure}

\begin{figure}
\centering
\includegraphics[width=125mm,height=100mm]{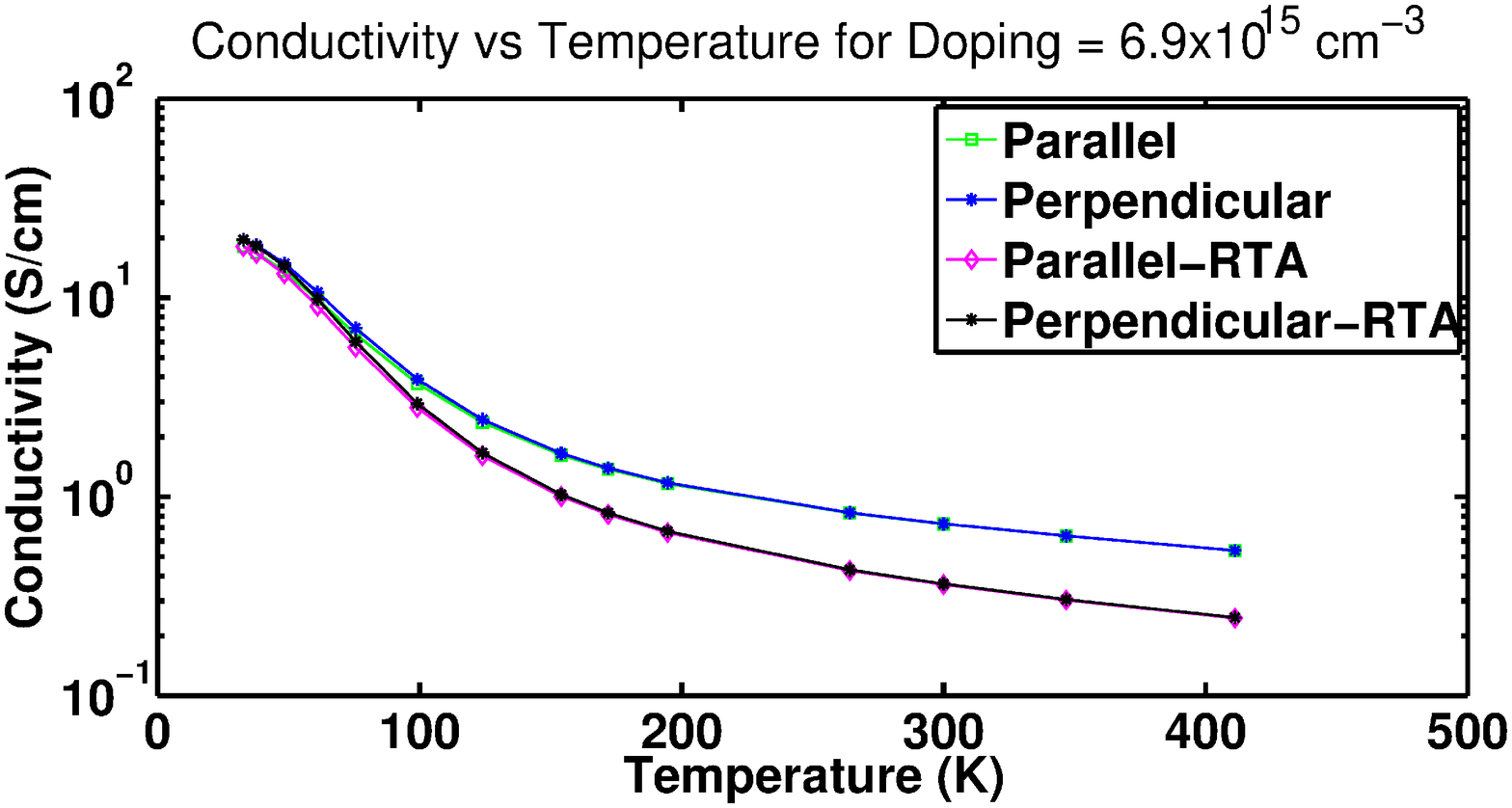}
\caption{Calculated conductivity with temperature variation for CdS at Doping  $6.9 \times 10^{15} cm^{-3}$ \cite{crandall}}
\label{c-t}
\end{figure}

\begin{figure}
\centering
\includegraphics[width=125mm,height=100mm]{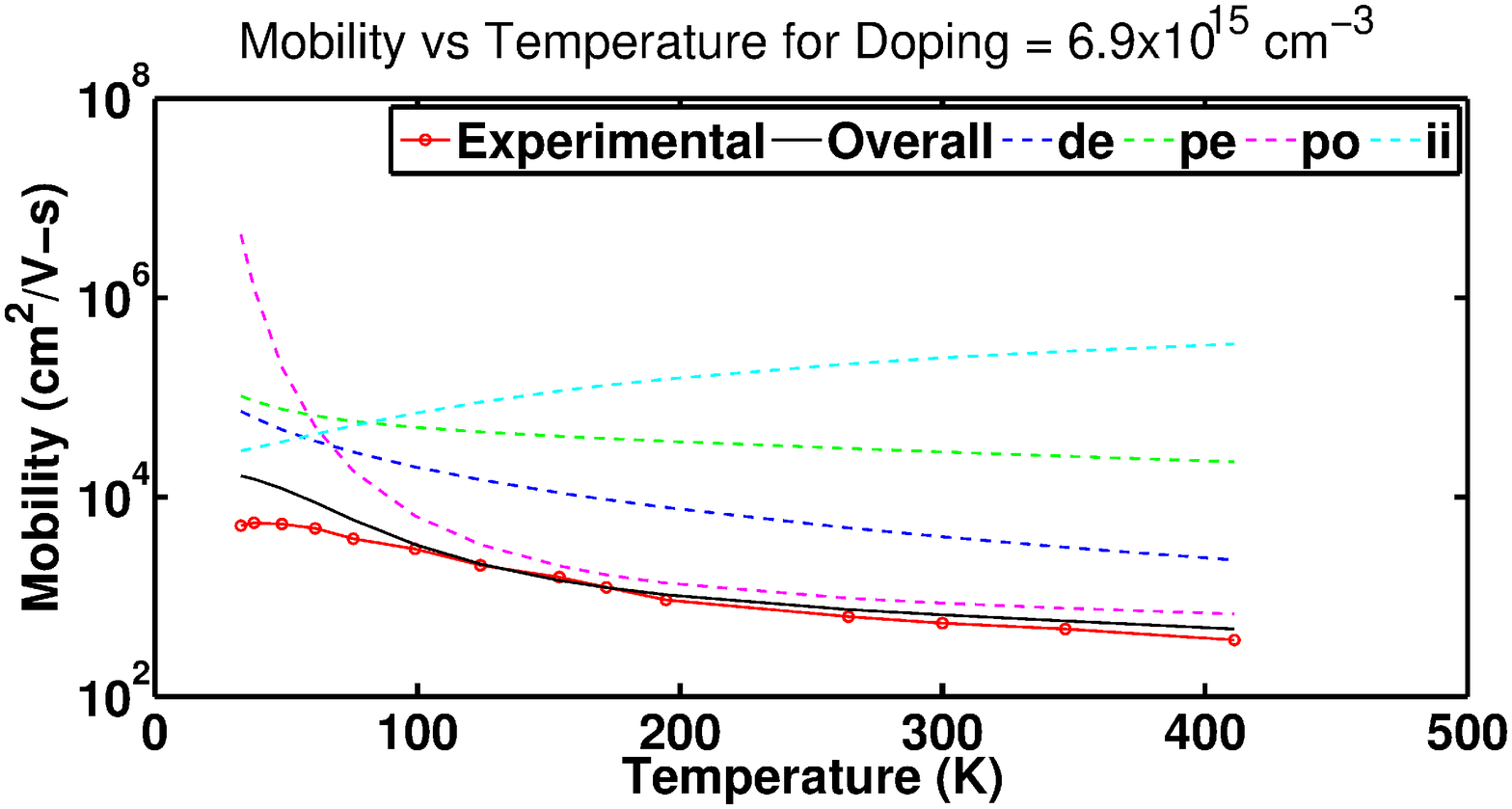}
\caption{Contribution of mobility from different scattering mechanisms at doping $6.9 \times 10^{15} cm^{-3}$ \cite{crandall}}
\label{m-all}
\end{figure}

\begin{figure}
\centering
\includegraphics[width=125mm,height=100mm]{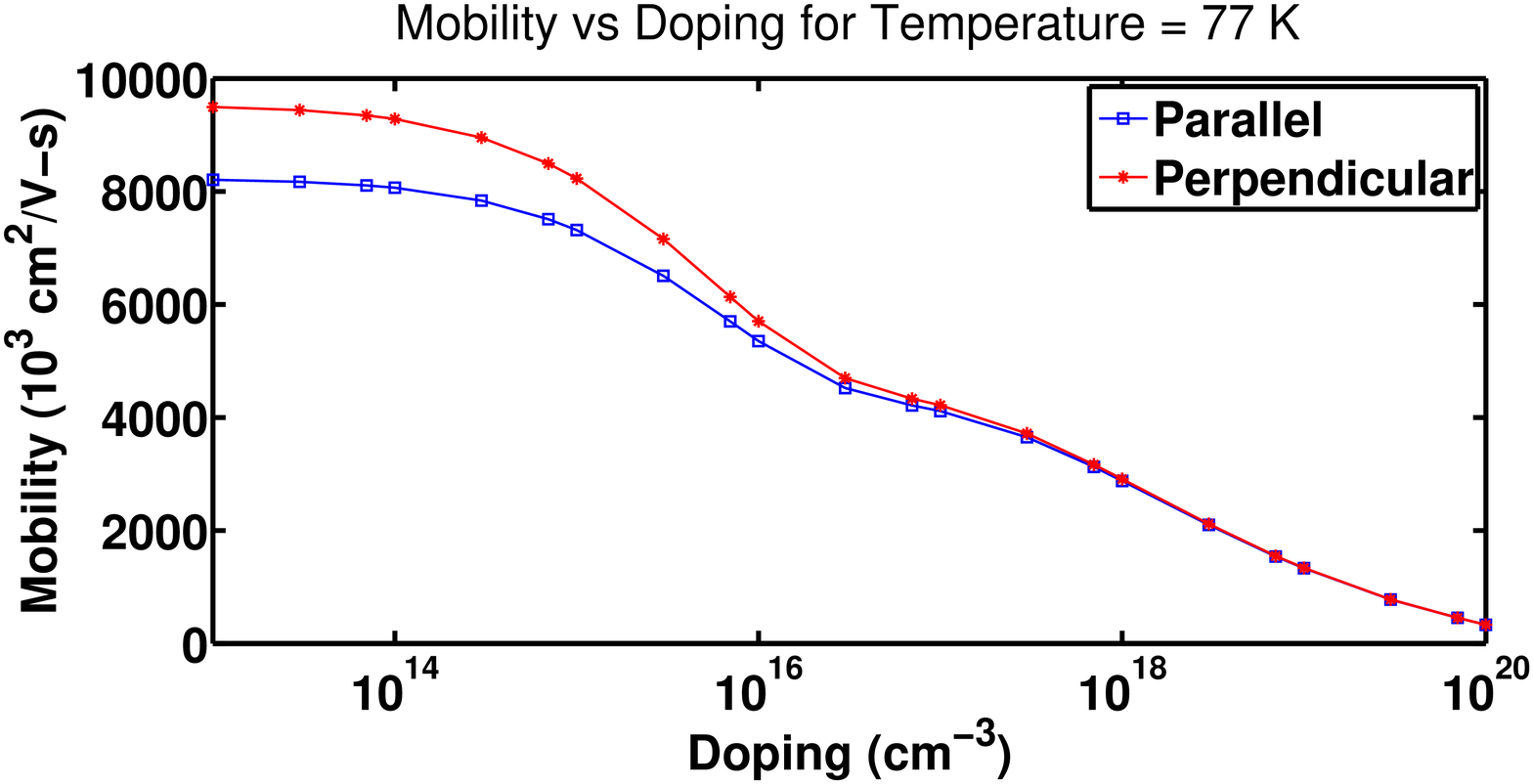}
\caption{Calculated mobility for different doping concentration at 77 K}
\label{m-d-77}
\end{figure}

\begin{figure}
\centering
\includegraphics[width=125mm,height=100mm]{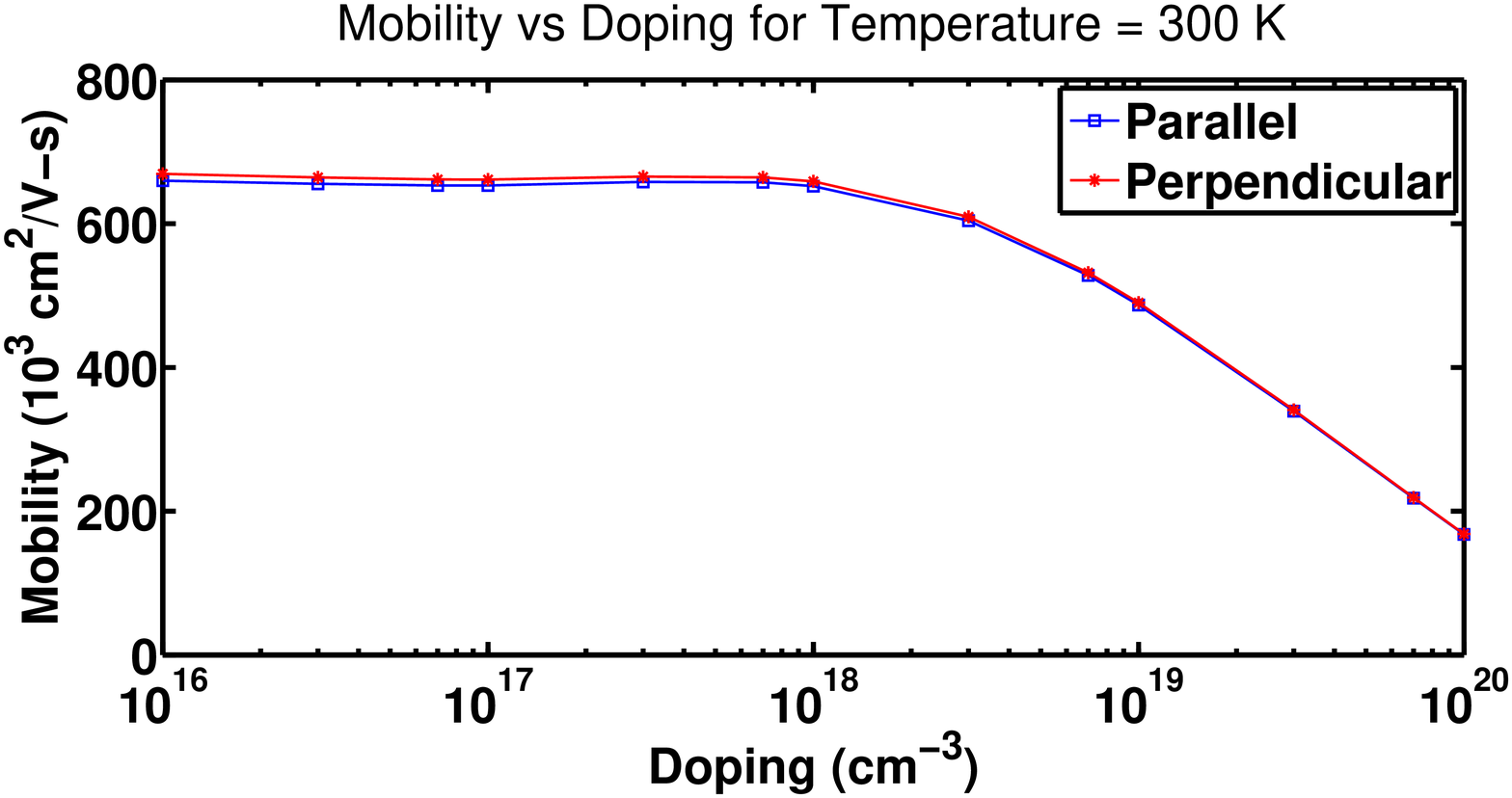}
\caption{Calculated mobility for different doping concentration at 300 K}
\label{m-d-300}
\end{figure}

\begin{figure}
\centering
\includegraphics[width=125mm,height=100mm]{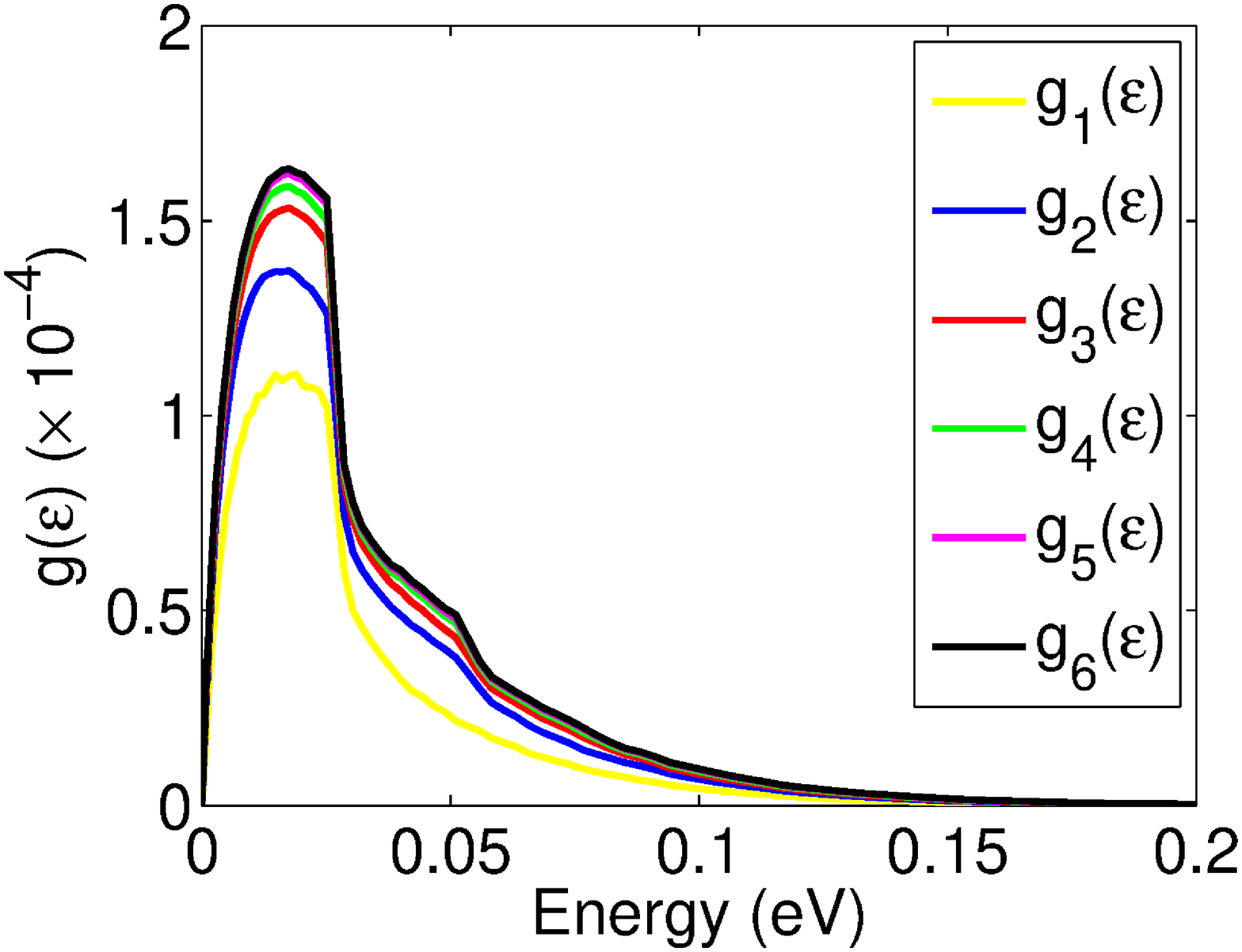}
\caption{The evolution of the perturbation part of the distribution function (Eq.\ref{pert}) with respect to different iterations.}
\label{g-E}
\end{figure}

Fig \ref{scattering_13} and \ref{scattering_15} shows scattering rate as a function energy for different doping concentration $N_D = 1 \times 10^{13} cm^{-3}$ and $N_D = 1 \times 10^{15} cm^{-3}$ at temperatures 30 K and 300 K. Piezoelectric scattering is considered to be the significant scattering mechanism at lower temperature and lower doping concentration. At 30 K average energies of carriers is $\frac{3}{2} k_B T = 0.0038 eV$, so most of carriers lies in the low energy region. For CdS at lower temperature the piezoelectric scattering is most dominant scattering  mechanism as in fig \ref{scattering_13}(a) for doping $N_D = 1 \times 10^{13} cm^{-3}$ while for doping $N_D = 1 \times 10^{15} cm^{-3}$ ionized impurity and piezoelectric scattering are most dominant scattering mechanisms as in fig \ref{scattering_15}(b). While at room temperature for both doping concentrations polar optical phonon scattering is the most dominant scattering mechanism for CdS as it is clear in both figures Fig. \ref{scattering_13}(b) and fig \ref{scattering_15}(b). So from room temperature to the higher temperature region POP scattering is most dominant one. In both Fig. \ref{scattering_13} and Fig. \ref{scattering_15} there is a sudden change in POP scattering rate after a particular energy, this is due to the fact that if an electron energy is less than POP energy $\hbar \omega_{po} =  0.0266 eV $, the electron can scatter only by the absorption of optical phonons and if the electron energy is higher than energy $\hbar \omega_{po}$, then it can scatter through both absorption and emission of phonons. \\
We have calculated mobility and conductivity of CdS using \verb|AMMCR| module for different temperature and doping concentrations. Fig. \ref{m-t} shows mobility and fig \ref{c-t} show conductivity as a function of temperature for doping $6.9 \times 10^{15} cm^{-3}$ \cite{crandall}. Fig \ref{m-t} shows good qualitative and quantitative agreement between experimental and calculated curves. 
Fig. \ref{m-t} and fig \ref{c-t} shows the results obtained with both Rode algorithm and RTA. Since piezoelectric constant is different in parallel and perpendicular to c-axis of CdS crystal, so there are two theoretical curves shown for both Rode and RTA. At lower temperature there is a deviation between experimental and theoretical curves, this is due to the presence of neutral impurities in the sample. Rode curve has an average relative error of 10.93 \% and RTA curve has an average error of 37.06 \% above 90 K with parallel piezoelectric coefficient. So, Rode results are much better than RTA results. This is due to POP scattering, since POP is inelastic as well as  an-isotropic scattering mechanism, so it modulated electron energy and it is most dominant scattering mechanism around room temperature to higher temperature region, so RTA is inappropriate for it. At a lower temperature, both RTA and Rode show about the same mobility. This is due to the fact that POP scattering is weak at a lower temperature. \\
As already mentioned, most of the compound semiconductors are polar in nature. In a polar semiconductor, usually polar scattering is the most dominant scattering mechanism.Therefore, \verb|AMMCR| will serve a better purpose in comparison to the RTA based codes such as Boltztrap.
Fig. \ref{m-all} shows the contribution of mobility from different scattering mechanisms. Above 70 K polar optical phonon scattering is most dominant scattering mechanism, below 55 K ionized impurity scattering is most dominant scattering mechanism and in the intermediate region acoustic deformation scattering is most dominant scattering. Fig. \ref{m-d-77} and fig. \ref{m-d-300} shows mobility variation with doping for temperature 77 K and 300 K respectively by assuming a compensation ratio of unity. The curves shows decrease in mobility with increasing doping concentration, this is due to increase in number of ionized centers with increasing doping. In the Fig.\ref{g-E}, we show the convergence of the non-equilibrium part of the distribution function. It can be seen that the convergence is achieved after a few iterations.

\section{Conclusion}
We presented a module \verb|AMMCR| to compute mobility and conductivity of semiconductor materials. The module is written in C++ and is currently interfaced with VASP. We have tested the code for CdS and obtained good agreement with the experimental results.\\
\section{Acknowledgments}
The authors acknowledge funding from Indo-Korea Science and Technology Center (IKST). We thank the Computing infrastructure from the Center of Excellence in Nanoelectronics, IIT Bombay and IKST for providing the computational facility.
\section{Appendix}
For zinc blende structure piezoelectric coefficient is given by \cite{rode3}
\begin{equation}
\ P^2 = h_{14}^2\frac{[(\frac{12}{c_l})+(\frac{16}{c_t})]}{35} 
\label{pzcoeff}
\end{equation}
where $h_{14}$ is one element of piezoelectric stress tensor and  $c_l$, $c_t$ are the spherically averaged elastic constant for longitudinal and transverse modes respectively and are given by \cite{rode3} equations
\begin{equation}
\ c_l = (3c_{11} + 2c_{12} + 4c_{44})/5 
\label{cl}
\end{equation}

\begin{equation}
\ c_t = (c_{11} - c_{12} + 3c_{44})/5 
\label{ct}
\end{equation}
where $c_{11}$, $c_{12}$ and $c_{44}$ are three independent elastic constants.
For wurtzite structure we use piezoelectric coefficients $P_\parallel$ and $P_\perp$ for mobility measured with electric field parallel and perpendicular to the c axis of the crystal. For wurtzite structure piezoelectric coefficients $P_\parallel$ and $P_\perp$ are given by \cite{rode3}
\begin{equation}
\ P_{\parallel}^2 = 2 \epsilon_0 \frac{(21 h_{15}^2 + 18 h_{15} h_{x } + 5 h_x^2)}{105c_t} + \epsilon_0 \frac{(63 h_{33}^2 - 36 h_{33} h_x + 8 h_x^2)}{105c_l}  
\label{pzcoeff_para}
\end{equation}

\begin{equation}
\ P_{\perp}^2 = 4 \epsilon_0 \frac{(21 h_{15}^2 + 6 h_{15} h_{x } + h_x^2)}{105c_t} + \epsilon_0 \frac{(21 h_{33}^2 - 24 h_{33} h_x + 8 h_x^2)}{105c_l}  
\label{pzcoeff_perp}
\end{equation}

\begin{equation}
\ h_x = h_{33} - h_{31} - 2h_{15}
\label{hx}
\end{equation}
where $h_{15}$, $h_{31}$ and $h_{33}$ are the three independent elements of the piezoelectric stress tensor of wurtzite structure and $c_l$ and $c_t$ are spherically averaged elastic constant, there are given by equations \cite{rode3}
\begin{equation}
\ c_l = (8c_{11} + 4c_{13} + 3c_{33} + 8c_{44})/15 
\label{cl_w}
\end{equation}
\begin{equation}
\ c_t = (2c_{11} - 4c_{13} + 2c_{33} + 7c_{44})/15 
\label{ct_w}
\end{equation}
where  $c_{11}$, $c_{13}$, $c_{33}$ and $c_{44}$ are elastic constants.

\section*{References}

\bibliographystyle{elsarticle-num}
%\bibliography{ref}

%------------------------------------------ figures and tables ----------------------------------
\newpage 

\clearpage
\newpage
\clearpage
\newpage
\clearpage
\newpage
\clearpage
\newpage
\clearpage
\newpage
\clearpage
\newpage
\clearpage
\newpage
\clearpage
\newpage
\clearpage
\newpage
%%-----------------------------------------------------------------------------------------------

\end{document}